\documentclass[fleqn,10pt,twocolumn]{wlscirep}
\usepackage{placeins}	
\usepackage{textcomp}
\usepackage{hyperref}
\usepackage{float}
\usepackage{parskip}
\emergencystretch=\maxdimen
\hypersetup{
    colorlinks = false,
}

\title{
Colors from correlated disordered photonic systems - can we outperform nature?
}

\author[1]{Gianni Jacucci}
\author[1,*]{Silvia Vignolini}
\author[1,*]{Lukas Schertel}

\affil[1]{University of Cambridge, Department of Chemistry, Cambridge, CB2 1EW, United Kingdom}
\affil[*]{Email: sv319@cam.ac.uk, ls849@cam.ac.uk}

\begin{abstract}
Living organisms have developed a wide range of appearances from iridescent to matt textures ~\cite{Vukusic2003, Johansen2017}.  Interestingly, angular independent structural colors, where isotropy in the scattering structure is present, only produce coloration in the blue wavelength region of the visible spectrum. One might, therefore, wonder if such observation is a limitation of the architecture of the palette of materials available in nature. 
Here, by exploiting numerical modeling, we discuss the origin of isotropic structural colors without restriction to a specific light scattering regime. We show that high color purity and color saturation cannot be reached in isotropic short-range order structures for red hues. This conclusion holds even in the case of advanced scatterer morphologies, such as core-shell particles or inverse photonic glasses - explaining recent experimental findings reporting very poor performances of visual appearance for such systems ~\cite{Kim2017, Schertel2019}.
\end{abstract}

\begin{document}
\maketitle
\noindent
\section{Introduction}
Structural colors are the results of constructive interference of light scattered by nanostructured, non-absorbing media. In contrast to color by pigmentation, where the color results from wavelength-selective absorption, structural colors spanning across the entire visible spectrum can be obtained using only one material. 
The possibility of exploiting such photonic coloration to replace traditional pigments catalyzed the efforts of several research groups ~\cite{Kolle2010, Forster2010, Magkiriadou:12, Vogel2015, Liang2018, Goerlitzer2018, Schertel2019a}. Structural colors come with various advantages compared to conventional pigments: (i) structural pigments do not bleach, as the color formation is defined by the architecture instead of the composition; (ii) they can be constituted of environmental-friendly materials ~\cite{Wang2018, Frka-Petesic2019}; (iii) they achieve unconventional color effects, from vivid metallic to isotropic optical response.

\begin{figure*}[h]
    \centering
    \includegraphics[width=.75\linewidth]{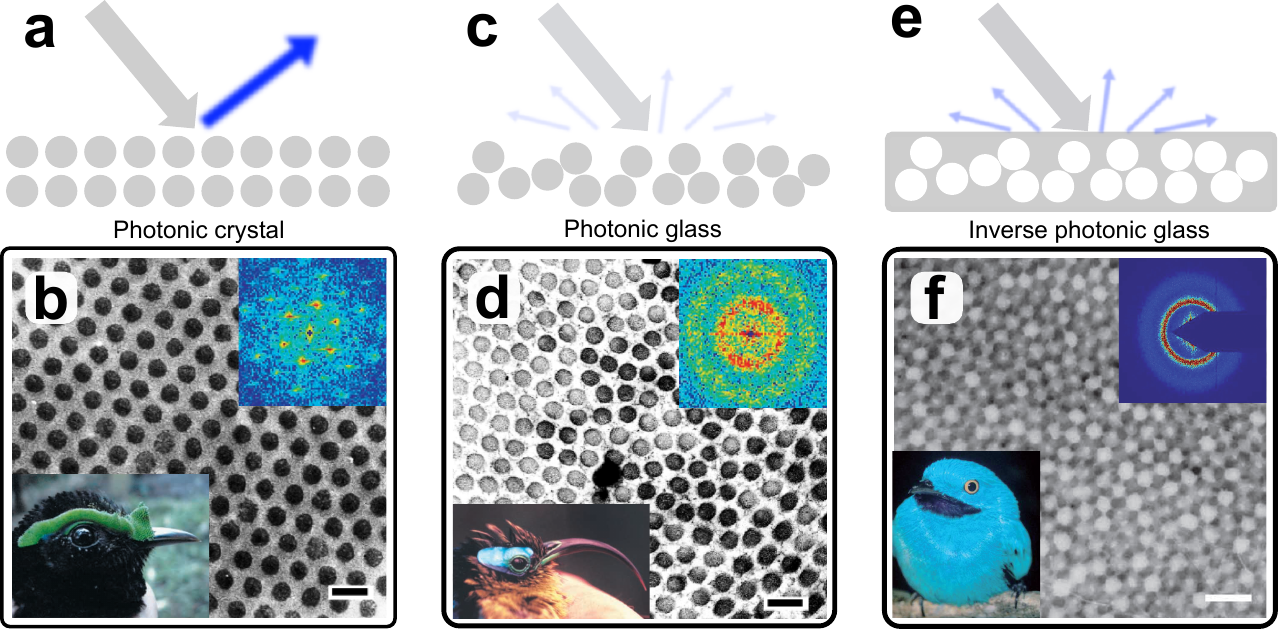}
    \caption{\textbf{Structural color in natural systems.}
    Sketch of light scattered by a) a photonic crystal, c) a photonic glass and e) an inverse photonic glass. b) Transmission
electron micrograph of collagen arrays from structurally colored facial caruncles of male asities \textit{Philepitta castanea} (lower left) and its 2D Fourier power spectra (upper right). d) TEM of collagen arrays from structurally colored facial caruncles of male asities \textit{Neodrepanis coruscans} (lower left) and its 2D Fourier power spectra (upper right)~\cite{Prum1999a}. b), d) Adapted with permission,~\cite{Prum1999a} Copyright 2020, Company of Biologists. f) EM image of sphere-type keratin and air nanostructure from back contour feather barbs of \textit{Cotinga maynana} (lower left) and corresponding small-angle X-ray scattering data (upper right). Adapted from Ref.~\cite{Dufresne2009} with permission from The Royal Society of Chemistry.
}
    \label{fig1}
\end{figure*}

Isotropic structural color is especially appealing when it comes to replacing traditional pigmentation as they preserve an angular independent appearance. Therefore, several methods were developed to produce hierarchical  \cite{Vogel2015,Song2018,Chan2019},  or short-range order structures with angular-independent colors \cite{Forster2010,Schertel2019, Kim2017}. 
Such studies of short-range order structures, often referred to as photonic glasses (PGs), found inspiration in analogous two-dimensional structures exploited in birds (\autoref{fig1}) ~\cite{Prum1998,Prum1999a,Prum2003,Saranathan2012}.  However, in nature, such PGs have been reported only to produce blue colors, as green and red coloration are usually achieved with long-range ordered structures or using pigmentation~\cite{Kinoshita2008,Sakoda2001}.

In the following, we investigate the limitations of color creation
by scattering from disordered, short-range correlated structures. We question whether a saturated, angular independent, color response can be achieved only in the UV-blue spectral region or whether nature has missed a trick and developed other mechanisms for red hues. In fact, while artificial PGs with red hues have been reported, their optical properties
are rather poor - their color saturation and purity are limited, holding these materials back from applications. These limitations of red isotropic structural color have been attributed to single-particle resonances ~\cite{Magkiriadou2014}. However, this conclusion ignored multiple scattering and coupling effects between particles, which are known to play a major role in PGs ~\cite{Schertel2019,Hwang2019}.
Here, we exploit a numerical approach that provides direct access to the reflection spectrum of an arbitrary structure and allows us to investigate intermediate scattering regimes, i.e., in between single scattering and diffusive behavior, without further assumptions.
With this method, we show that nature's solution is actually optimal for blue production only: both high and low refractive index contrast PGs show poor color purity in the red. 
We demonstrate the difficulty in achieving isotropic structural color in the red spectral region with high saturation and purity even for advanced morphologies of the scatterers in both direct and inverse PGs.

\section{Results}
\subsection{Direct vs inverse photonic glasses}
Photonic glasses with tailored form and structure factors were generated using a recently developed numerical algorithm, see details in Reference \cite{Jacucci2019} and Supplementary Section III. The optical properties of the generated structures were then calculated using a \textit{Finite Difference Time Domain }(FDTD) method. We limited our calculations to two-dimensional structures, as the same general considerations can be extended to the three-dimensional case.

\begin{figure}[t]
    \centering
    \includegraphics[width=\linewidth, page=2]{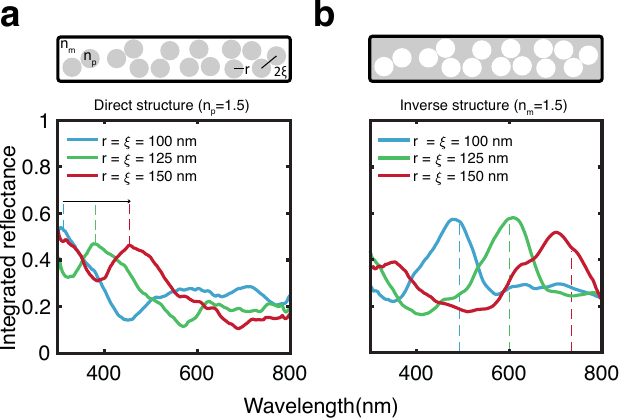}
    \caption{\textbf{Simulated optical response for direct and inverse photonic glasses.} Inverse PGs, where scattering is dominated by structural correlations, exhibit a spectral peak whose position can be tuned over the whole visible range controlling the size of the scatterers. Dotted lines indicate: the position of the first Mie resonance in direct PGs and the Bragg's law prediction in inverse PGs, respectively.
All the simulated structures have a thickness of 3 $\mu m$ and  ff = 0.5.}
    \label{fig2}
\end{figure}

In the absence of absorption, scattering in PGs emerges from the interplay between: (i) single particle properties as size, shape and refractive index; (ii) ensemble properties as filling fraction and structural correlations.
As depicted in \autoref{fig2}a, for the case of direct PGs the reflection is dominated by Mie resonances, determined by the scatterer properties. The reflected color can be, therefore, tuned in the visible by changing the scatterer dimensions. However, as the particle size increases the Mie resonance peak red-shift and a second peak appears in the blue part of the spectrum, corresponding to a higher-order resonant mode (Figure S1).
In contrast, light scattering in inverse PGs is dominated by structural correlations (\autoref{fig2}b). The reflection peak, whose position corresponds well to Bragg's law predictions, is more pronounced than in direct structures. Moreover, when the structural peak is shifted at red wavelengths the form-factor resonance at short wavelengths is less marked than in \autoref{fig2}a. 
The occurrence of a separated peak in the visible spectrum showcases that using inverse PGs is an effective strategy to minimize the form-factor weight on the overall optical response of a system in favor of structural contributions.

\subsection{The role of refractive index (contrast)}
The dependency of isotropic structural coloration on the refractive index is shown in Figure S1 and Figure S2 for direct and inverse PGs, respectively. Changing the refractive index affects the interplay between form- and structure- factor contributions. High refractive index systems are dominated by form-factor resonances preventing to reach good color purity in the red spectral region, both for direct and inverse PGs. 
For direct systems, even when the refractive index contrast is low, form-factor resonances lead to an enhanced reflection in the short wavelength side of the structural peak - similarly to \autoref{fig2}a. In contrast, for the case of inverse PGs, we observe that the structure-factor forms a well separated peak in the visible spectrum, even in the red wavelength region. This allows us to conclude that low refractive index, inverse PGs can outperform their direct counterpart in terms of color purity and saturation. 

Reducing the refractive index contrast between the scattering matrix ($n_m$) and the scattering centers ($n_p$) can further favor structural contributions. 
\autoref{fig3new}a shows that increasing $n_p$ results in a broadband reduction of the reflectance and a red-shift of the structural peak. Moreover, the structural peak decreases in width and has a higher intensity compared to its background, leading to better color purity. 
The reduction of the refractive index contrast lowers the role of multiple scattering, which is anyway present in disordered systems. This limits the isotropic structural colors to a light propagation regime in between diffusive scattering and ballistic transport. 
The multiple scattering becomes dominant when the sample thickness is increased, leading to a broadband unsaturated response (Figure S3a). 

To further investigate the advantages and limitations of isotropic structural coloration in direct and inverse PGs, we studied their dependency on different ensemble parameters.  Figure S3b shows the optical response of inverse PGs is more robust to variations in scatterer size distribution than the direct PGs. While for direct PGs polydispersity averages out the Mie resonances, in inverse PG polydispersity only slightly affects the structure-factor resonance peak.  
In inverse PGs, the polydispersity helps to reduce the Mie resonance intensity in the blue region and broadens the structural peak at long wavelengths. 
Similarly, reducing the filling fraction in inverse PGs leads to a red-shift of the structural peak, due to an increase of the effective refractive index of the system. In parallel, the relative intensity of the structural peak decreases compared to the Mie resonance at short wavelengths (Figure S3c). 
The result of these effects on the overall appearance of the system is further studied in \autoref{fig4}.

\subsection{Advanced scatterer morphologies}
\begin{figure}[H]
    \centering
    \includegraphics[width=\linewidth, page=4]{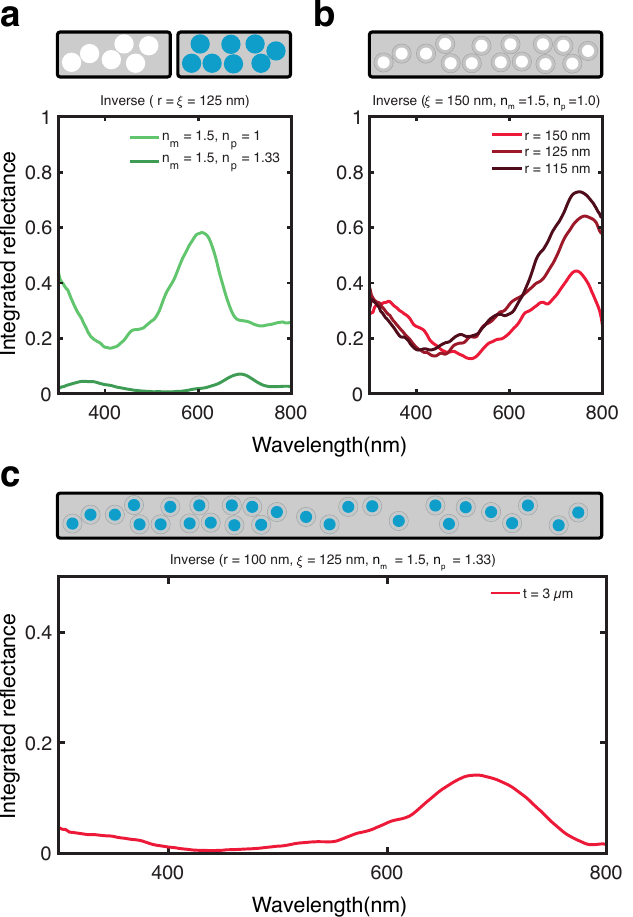}
    \caption{\textbf{Simulated optical response for inverse photonic glasses with advanced designs.} 
a) Effect of the refractive index of the scatters; decreasing the refractive index contrast between scatterers and matrix  leads to a broadband decrease of the reflectance, increasing the color purity.
b)  Effect of core-shell correlation; decreasing the size of the scatterers ($r$, core) maintaining a fixed center-to-center correlation distance ($\xi$, shell) leads to a red-shifted peak, due to an increase of the effective refractive index of the system.
All the simulated structures have a thickness of 3 $\mu m$ and  ff = 0.5.}
    \label{fig3new}
\end{figure}

Our observations remain valid also for complex scatterer geometries. Previous works introduced the idea of using core-shell particles to disentangle the contributions of form- and structure-factor and achieve a separated peak in the long wavelength spectral region \cite{Magkiriadou:12,Park2014,Magkiriadou2014, Shang2018}.
\autoref{fig3new}b shows that reducing the scattering center (core) size while keeping the structural correlation length leads to an increase of the intensity and width of the long wavelength (structural) peak. At the same time the short wavelength contribution from Mie resonances shifts further away into the UV.
In \autoref{fig3new}a we showed that a lowered refractive index contrast can suppress multiple scattering, while a separation of the form- and structure-factor contributions is possible via core-shell particles, see \autoref{fig3new}b. In \autoref{fig3new}c both approaches are combined, to achieve higher color purity and saturation by a well separated peak in the long wavelength part of the visible spectrum.

\subsection{Color saturation and purity}
\begin{figure*}[h]
    \centering
    \includegraphics[width=\textwidth]{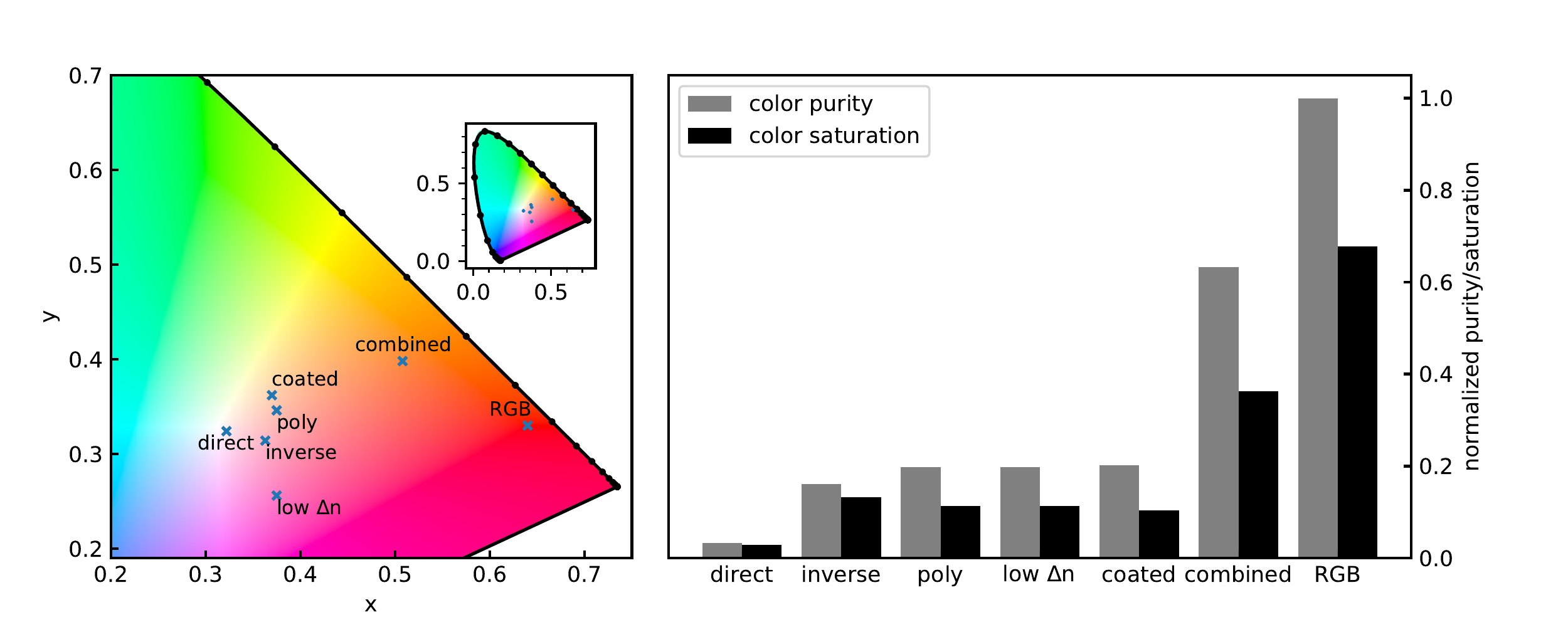}
    \caption{\textbf{Color purity and saturation.} a) Zoom-in to the CIE chromaticity diagram (inset). The xy values of the isotropic structural colors for various systems are displayed (see SI for values): (direct) the ECPA multiple scattering model (see~\cite{Schertel2019} Fig S3, $r=128$~nm), (inverse) an inverse PG with an isolated peak in the high wavelength region (red curve in \autoref{fig2}b), (poly) the same structure with $20\%$ polydispersity (dark red in Figure S3c), (low $\Delta$n) an inverse PG with lowered refractive index contrast (dark green in \autoref{fig3new}a), (coated) a core-shell inverse PG ($r=115$~nm in \autoref{fig3new}b) and (combined) a combined coated and lowered refractive index approach (\autoref{fig3new}c). For comparison RGB red color is displayed (RGB). b) The corresponding saturation and purity values are shown.}
    \label{fig4}
\end{figure*}

To better quantify and assess the results in terms of color purity and saturation, the reflectance spectra of direct, inverse and core-shell PGs are converted to color hues. The details of the spectrum to color conversion are reported in the Methods section. In \autoref{fig4}a the different systems for red color hues are plotted in the CIE color space chromaticity diagram (see
Table S1 for the xy-values). In \autoref{fig4}b the corresponding purity and saturation values are calculated. Interestingly, all the inverse PGs show a higher color purity and saturation value than the direct PGs red hues reported in Reference~\cite{Schertel2019}. However, both the core-shell inverse structures as well as the low refractive index contrast (e.g. polymer filled with water) do not lead to a significant improvement compared to the standard inverse PG. Combining both approaches leads to improved purity and saturation values. Still, the achievable color purity and saturation values remain far from the ideal (RGB) red color.

\section{Discussion}
In summary, we demonstrated that photonic glasses have intrinsic limitations in achieving saturated red hues - reinforcing the hypothesis that isotropic structural coloration, while it can be easily achieved in the UV-blue spectral region, is challenging to obtain for larger wavelengths. 
In detail, we showed that inverse PGs increase color saturation and purity compared to their direct counterpart. This result is a consequence of an optical response where the structure factor contribution dominates over the form-factor scattering resonances. 

While combing different scattering approaches can help improving color purity and saturation in photonic glasses, they might be more challenging to realize in natural systems than simply combing different mechanisms, such as adding absorption.
Once absorption is added, more pure and saturated colors can be achieved~\cite{Goerlitzer2018, Kim2019}. However in synthetic systems, especially in the case of metallic nanoparticles ~\cite{Lee2019}, one might question the need of adding a structural component, as it plays a redundant role in the overall visual response. In this context, we believe that approaches aimed at designing the scattering elements, making use of the interplay between absorption and scattering via Kramer-Kronig relations might be more suitable to increase scattering in the red spectral region ~\cite{Wilts17}.
Although most efforts to fabricate systems exhibiting isotropic structural colors focus on the self-assembly of colloidal particles \cite{Goerlitzer2018}, other promising approaches have recently been explored combining the purity and saturation advantages of coloration from ordered structures with the morphological isotropy of spherical systems. In this context, blockcopolymers represent a promising candidate ~\cite{Song2019}. Similarly, introducing disorder to ordered multilayer films gives rise to angle-independent structural coloration ~\cite{Song2018, Chan2019}.
In conclusion, as convincing  absorption-free, angle independent, pure, and saturated red hues have not been obtained, we believe there is still space for further theoretical and experimental investigations to come up with strategies to produce viable alternatives to red pigments.

\section*{Methods}
\subsection*{\label{sec:Numerics} Numerical calculations}
To generate the, two-dimensional, disordered structures we used a inverse-desing algorithm consisting of two steps: \cite{Jacucci2019} First, hard (non-overlapping) particles are added using a random sequential approach until the desired filling fraction was reached; Second, the position of these particles are gradually changed in order to minimize the difference between the targeted S(q) and the one of the structure.
This code is available from G. J. upon request.

To simulate the optical properities of the generated designed we used LUMERICAL 8.22 (Lumerical Solutions Inc., Vancouver, BC, Canada), a commercial-grade software using the finite-difference time-domain (FDTD) method. Periodic boundaries conditions in the lateral direction, i.e., perpendicular to the incoming beam, Y and perfect matching layer (PML) boundaries in the X direction were used in all the calculations. The excitation source was set as a plane wave. The simulations were performed in a purely 2D geometry and their numerical stability/convergence was ensured by choosing an adequate simulation time and boundary conditions (assuring that the electric field in the structure decayed before the end of the calculation and that all the excitation light was either reflected or transmitted). 
Each of the presented curves was obtained averaging the optical simulations of seven different ensembles of particles with identical parameters.

\subsection*{\label{sec:PandS} Purity and saturation of color}
To quantify color purity and saturation, reflection spectra are converted in color space coordinates assuming a standard observer (CIE 1931 2$^\circ$) and a standard illuminant (Daylight $D_{65}$). \cite{Schertel2019,Jacucci2019}

In particular, color purity can be calculated from: 
\begin{equation}
p_{e}=\sqrt{\frac{\left(x-x_{n}\right)^{2}+\left(y-y_{n}\right)^{2}}{\left(x_{I}-x_{n}\right)^{2}+\left(y_{I}-y_{n}\right)^{2}}} .
\end{equation}
where $x$, $y$ are the chromaticity coordinates in the xyY color space, with $x_{n}=0.31271$, $y_{n}=0.3290$ and $x_{I}=0.64$, $y_{I}=0.33$ being the coordinates of the white point and RGB red, respectively. color saturation is calculated as:
\begin{equation}
s_{a b}=\frac{C_{a b}^{*}}{L^{*}}=\frac{\sqrt{a^{* 2}+b^{* 2}}}{L^{*}},
\end{equation}
where, $L^{*}$ represents the lightness, and $a^{*}$, $b^{*}$ are the chromaticity coordinates in CIELAB space. 

\section*{Acknowledgements}
This work was supported in part by a BBSRC David Phillips Fellowship (BB/K014617/1), the European Research Council (ERC-2014-STG H2020 639088) to S.V and G.J and the Swiss National Science Foundation under project P2ZHP2\_183998 to L.S.

\section*{Data availability}
All data needed to evaluate the conclusions in the paper are present in the paper and/or the Supplementary Materials.

\section*{Conflict of interest}
The authors declare that they have no competing financial interests.

\bibliography{ms}

\end{document}


\maketitle
\noindent

\section{\label{sec:ECPA}Role of form factor and structure factor in direct photonic glasses}
The optical appearance of a material depends on the illumination source, the observer's receptors sensitivity spectrum and the reflection properties of the material. The latter is characterized by the reflection spectrum $R(\lambda)$. In the limit of non-absorbing, diffusive media it can be calculated via the transmittance of a slab shaped sample which is directly related to the optical density $\ell^{*}/L$, with $\ell^{*}$ the transport mean free path and $L$ the sample thickness~\cite{Garcia1992}. $\ell^{*}$ defines the length scale after which isotropy in the multiple scattering is reached and is inversely related to the turbidity of a multiple scattering sample.

For high density PG made of monodisperse spheres with radius $r$ and filling fraction $ff$ a recently developed scattering model can be used to calculate the scattering strength $\lambda/\ell^{*}$ (with $\lambda$ the wavelength of the incident light) from the anisotropy factor $\langle \cos \theta \rangle $ and scattering cross section~\cite{AubryandSchertel2017} $\sigma_{\mathrm{s}}$:
\begin{equation}
\frac{\lambda}{\ell^{*}}=  \frac{\lambda(1-\langle \cos \theta \rangle )3ff \sigma_{\mathrm{s}}}{4 \pi r^3}
\end{equation}
Expressing $\sigma_{\mathrm{s}}$ via the angular integral over the local scattering intensity $I(\theta)=F(\theta)S(\theta)$
\begin{equation}
\frac{\lambda}{\ell^{*}} = \frac{(1-\langle \cos \theta \rangle )3ff}{16 \pi^2} \left( \frac{\lambda}{r} \right)^3 \int_0^\pi F(\theta) S(\theta) \sin \theta \, \mathrm{d} \theta ,
\label{eq:scatteringStrength}
\end{equation}
shows that this calculation allows to disentangle the scattering contributions of the Mie form-factor $F(\theta)$~\cite{Bohren1998} and the structure-factor $S(\theta)$~\cite{Percus1958}. Note that in this formulation both $\langle \cos \theta \rangle $ and $\sigma_{\mathrm{s}}$ depend strongly on the choice of an effective refractive index of the material $n_\mathrm{eff}$~\cite{AubryandSchertel2017,Schertel2019a}. 
In this work, the energy coherent potential approximation (ECPA) refractive index is used as it accounts for near field coupling of the scatterers.~\cite{Busch1995} 
In summary, the reflection spectrum $R(\lambda)$ of a PG material can be calculated via $\ell^{*}$ and depends on $\lambda$, $r$, $ff$, the scatterer refractive index $n$ and the refractive index of the surrounding medium $n_m$.

\begin{figure*}[h]
    \includegraphics[width=\textwidth, page=1]{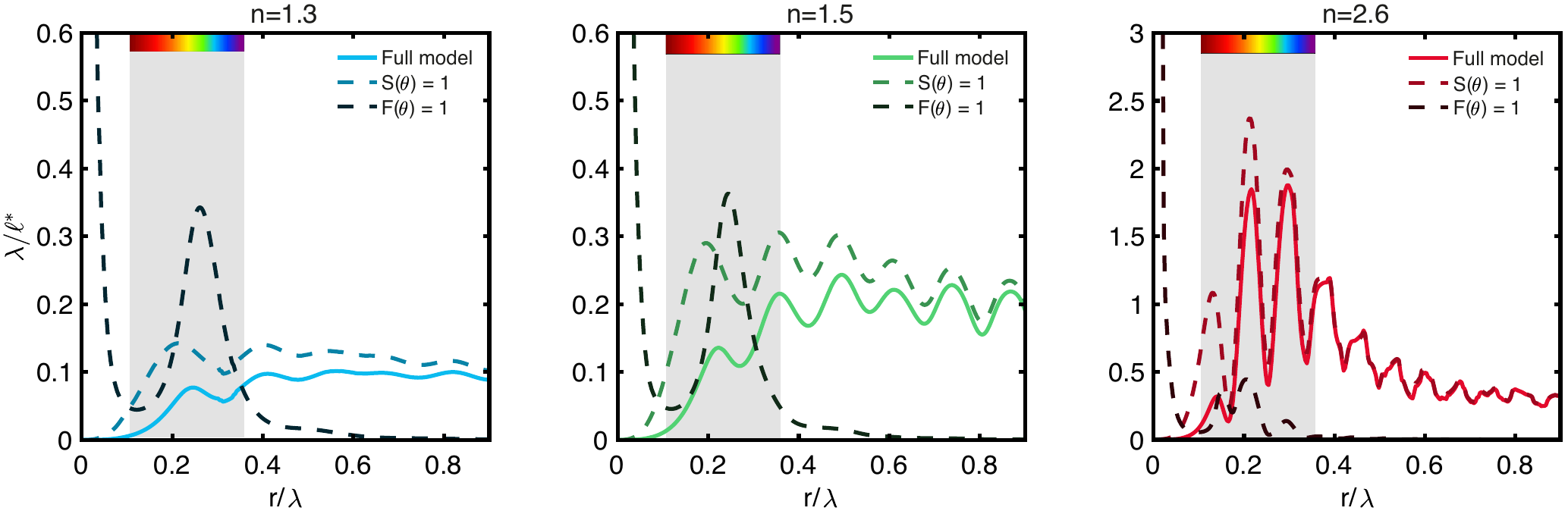}
    \caption{ECPA scattering model: scattering strength $\lambda/\ell^{*}$ with varying size ratio $r/\lambda$ calculated from \autoref{eq:scatteringStrength}, for Mie-resonance contributions only ($S(\theta)=1$) and for structural contributions only ($F(\theta)=1$) for three refractive indices of the scattering particles $n=1.3$, $n=1.6$ and $n=2.6$. For the surrounding medium a refractive index of $n_m=1.0$ was assumed. The visible wavelength range indicated by the grey shaded area refers to particles with r=130nm.
Models taking into account only structural contributions are scaled by a factor 0.01.}
    \label{fig:ECPA}
\end{figure*}


In Figure S1, we use this model to calculate the scattering strength for various refractive index contrasts: weakly scattering (e.g. index matched PG, $n=1.3$, blue solid line), polymer-like systems ($n=1.5$, green solid line) and highly scattering (e.g. titanium dioxide PG, $n=2.5$, red solid line).
We observe that the contribution from structural correlation (black dashed line) remains constant for all refractive indices while the contribution from Mie scattering (colored dashed lines) strongly scales with refractive index contrast. For high refractive index contrast ($n=2.6$) the scattering strength is dominated by the contribution from the form-factor, while in an intermediate regime ($n=1.5$) the position and amplitude of the scattering resonances is equally influenced by the structural correlation and the form-factor contribution, especially for the first multiple Mie-scattering resonance. For low refractive index contrast ($n=1.3$) the main spectral feature is strongly correlated with the position of the structural correlation.

\section{\label{sec:RI}Role of refractive index on the optical response of inverse photonic glasses}
Figure S2, generalizing what reported in Figure 2 of the main text, shows the importance of the refractive index on the optical response of inverse PGs. 
Indeed, as discussed in the main text, the structural color of PGs is the result of the balance of single-particle and structural contributions.
The spectral response of inverse PGs with n=2.60 is dominated by single-particle scattering; while for n=1.30 the structure factor contribution prevails.

\section{\label{sec:RI}Role of different ensemble parameters on the optical response of inverse photonic glasses}
Figure S3 shows how the optical response of PGs is effected by various ensemble parameters.
As discussed in further details in the main text, inverse PGs are more robust than direct PGs to changes of the polidispersity (Figure S3b). Moreover, when increasing the thickness the presence of multiple scattering leads to an increase of the incoherent background - therefore decreasing the purity and saturation of the optical response (Figure S3a).
Finally, decreasing the filling fraction of air inclusions leads to a red-shifted peak which is less pronounced than its counterpart at ff=0.5 (Figure S3c).

\begin{figure}
    \centering
    \includegraphics[width=0.5\linewidth]{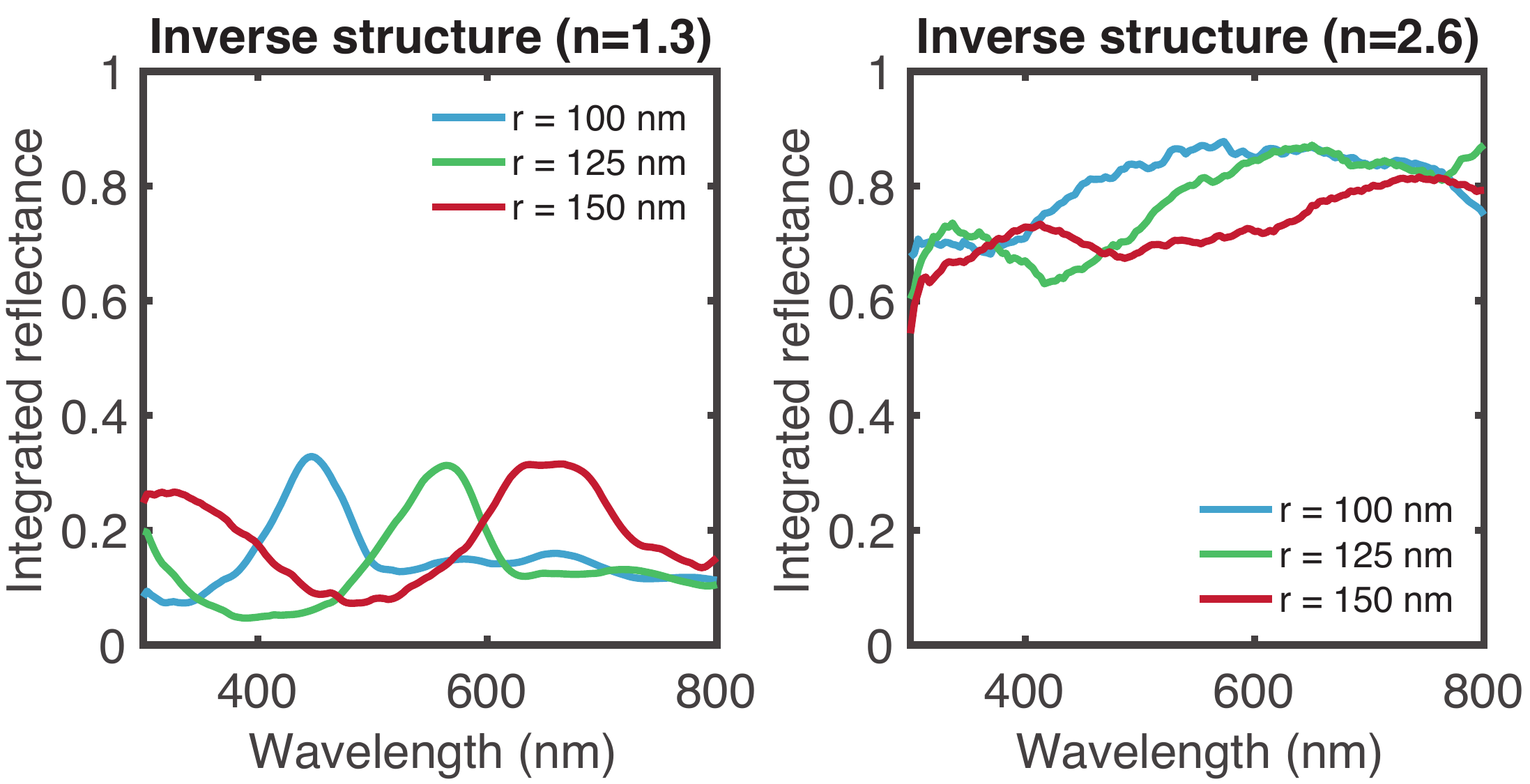}
    \caption{Simulated optical response for inverse photonic glasses with refractive index of n=1.3 and n=2.6, left and right panel, respectively. All the simulated structures have a thickness of 3 $\mu m$,  ff = 0.5.}
    \label{figS2}
\end{figure}

\begin{figure}[h]
    \centering
    \includegraphics[width=\linewidth, page=5]{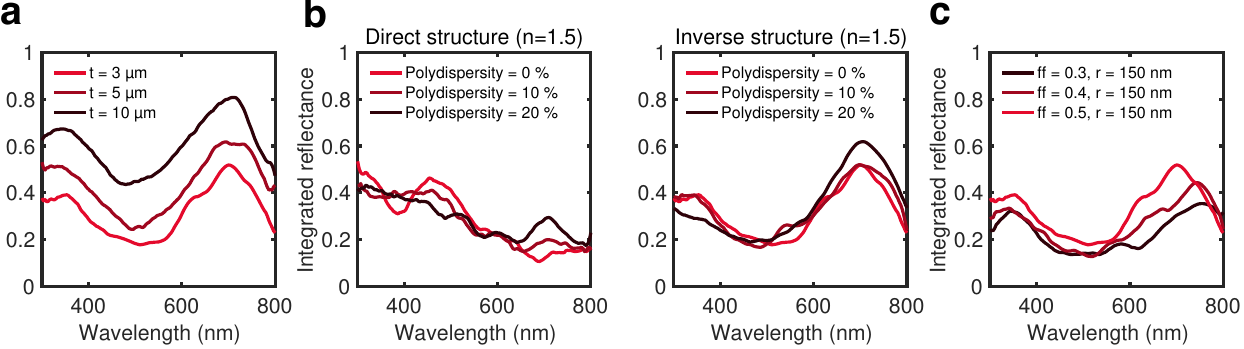}
    \caption{a) Effect of the thickness on the optical response of inverse photonic glasses; an increase in the thickness leads to a broadband increase of the reflectance, decreasing the colour purity.b) Simulated optical response for direct and inverse photonic glasses in function of particles' polidispersity. The spectral peak is more resistant to polidispersity in the case of inverse PGs, where it is mainly originating from structural contributions. 
c) Effect of the filling fraction on the optical response of inverse photonic glasses; a decrease in the filling fraction leads to a red-shifted peak, due to an increase of the effective refractive index of the system, with lower intensity.  
All the simulated structures have an average size of the scatterers of $r=150$ nm and, when not mentioned differently in the caption, a thickness of 3 $\mu m$,  ff = 0.5 and n=1.50.}
    \label{fig3}
\end{figure}

\section{\label{sec:xy-values} Chromaticity values displayed in Figure 4a}
\begin{tabular}{c|c|c|c}
   PGs strcuture & $x$ value & $y$ value & displayed spectrum \\ \hline
     direct & 0.321 & 0.324 & Fig SI.3 in ref.~\cite{Schertel2019} \\
     inverse & 0.363& 0.314& Fig 2b \\
     poly & 0.375& 0.346& Fig S3c \\
     low $\Delta$n & 0.375& 0.256& Fig 3a\\
     coated&0.37&0.362& Fig 3b\\
     combined&0.508&0.398& Fig 3c\\
     RGB&0.64&0.33& - \\
\end{tabular}

\newpage
\bibliography{supplement}